# PROPOSED INFORMATION SHARING SECURITY APPROACH FOR SECURITY PERSONNELS, VERTICAL INTEGRATION, SEMANTIC INTEROPERABILITY ARCHITECTURE AND FRAMEWORK FOR DIGITAL GOVERNMENT

Md.Headayetullah[1], G.K. Pradhan[2], Sanjay Biswas[3] , B. Puthal[4]

[1, 2, 3, 4] Department of Computer Science and Engineering, SOAU,
Bhubaneswar, India Phone: +91-9732062998
[1] headayetullahphd@gmail.com,
[2]gopa_pradhan@yahoo.com
[3]Drsanjay_Biswas@rediffmail.com
[4]bputhal@iter.ac.in

## *ABSTRACT*

*This paper mainly depicts the conceptual overview of vertical integration, semantic interoperability architecture such as Educational Sector Architectural Framework (ESAF) for New Zealand government and different interoperability framework solution for digital government. In this paper, we try to develop a secure information sharing approach for digital government to improve home land security. This approach is a role and cooperation based approach for security personnel of different government departments. In order to run any successful digital government of any country in the world, it is necessary to interact with their citizen and to share secure information via different network among the citizen or other government. Consequently, in order to smooth the progress of users to cooperate with and share information without darkness and flawlessly transversely different networks and databases universally, a safe and trusted information-sharing environment has been renowned as a very important requirement and to press forward homeland security endeavor. The key incentive following this research is to put up a secure and trusted information-sharing approach for government departments. This paper presents a proficient function and teamwork based information sharing approach for safe exchange of hush-hush and privileged information amid security personnels and government departments inside the national boundaries by means of public key cryptography. The expanded approach makes use of cryptographic hash function; public key cryptosystem and a unique and complex mapping function for securely swapping over secret information. Moreover, the projected approach facilitates privacy preserving information sharing with probable restrictions based on the rank of the security personnels. The projected function and collaboration based information sharing approach ensures protected and updated information sharing between security personnels and government intelligence departments to keep away from frightening activities. The investigational results exhibit the usefulness of the proposed information sharing approach. In a nutshell, this proposed security approach will work in together intra-organizational and inter-organizational department such as among security personnel and government intelligence department and this will make possible vertical information integration over and above it will progress interoperability approach for several recent digital government.*

# 1. INTRODUCTION

Government is a user of information technologies, donor of information based services and chief collector and provider of data and information [1]. Globally, now many governments facade challenges through usual way of renovation and so demand re-inventing the government systems thus as to reveal proficient and gainful services, information and knowledge via communication technologies [2]. A key step in re-inventing government is through nurturing digital government, which is nothing but the usage of general applications of information and communication technology that grips each transaction of government services [3]. Digital government is classically termed as the creation and transmittal of information and services inside government and among government and the public through a range of information and communication technologies. The impact of digital government varies largely across the world and is also well-known as e-government or virtual government [4]. E-government services are well-known in multifaceted architectural and technological circumstances [5]. Information age technologies pay for enormous prospects for a government to renovation its functions into the digital arena. Many government agencies have tightly engaged information technologies for renewing the government's vastly rambling service-centric information infrastructure by increasing information flow and the executive processes [6]. Information is a crucial facet of government's resources. So, currently, an urgent need to convince and approve larger flow of information is in demand along with data sharing among public agencies [7], [8]. The particular uprising in information resulted on organizations in the entire world to deeply rely upon huge numbers of databases to achieve their daily trade [9]. "Sharing information" is termed as the gathering and sharing of intelligence between two security divisions, or sharing innovative e-crime data, observations on these data, surveillance notes, scientific facts, commercial transaction data, and other. Information differs in the level of aspect, the quantity or type of data exchanged. Due to lack of standard methods for e-government information sharing, the means of sharing, at present are not consistently monitored, legitimated and recorded [10]. The information sharing environment is obscure and innovative resolutions and partnerships are crucial to collect shared benefits [11]. Moreover, the sharing is not constantly assured to be risk-free from risks that might grip illegal access, malicious alteration, and destruction of information or propaganda, computer intrusions, copyright infringement, privacy violations, human rights violations and more [10]. Since government departments are in need to share information within the same government and also across governments, the devising of an effective security appraise is essential. Tranquil, a department cannot aimlessly disclose its database to whichever another departments [12, 7]. A confined and trusted information-sharing environment is a prerequisite to enable users to communicate with and share information lacking difficulty and perfectly across a lot of distinct networks and databases unanimously. This means can considerably press forward the worth of overflowing functions, such as intelligence assembly and public wellbeing efforts [13, 14, 15, 16]. Guarantying security for its information systems, together with computers and networks, is a fundamental need for a digital government to function to the hope of its people. Information security is nothing but defending information and information systems from unauthorized access, use, disclosure, disruption, modification, or destruction. The key elements of information security encompass integrity, confidentiality, availability, authentication that has to be considered at various levels inside the hierarchy [17]. Production of a wide basement for information sharing desires trust among all information sharing partners. Insufficiency of trust leads to fears that shared information will not be secluded normally or used properly; and, that sharing will not continuously happen in both directions [18]. By using a safe and sound information sharing system, organizations can put in with promise in communities of trust since for this reason they have the controls so as to exactly direct their information accessing and usage. Let us deem a confined law enforcement officer at a usual traffic stop. Basic protocol utters that the officer request and confirm the individual's driving license and vehicle registration. Still, the officer might in addition check a wide range of other computer applications, such as colonization databases, terrorist watch lists, criminal information and intelligence repositories, and counter-drug intelligence databases that may be owned by exterior organizations, such as the U.S. Federal Bureau of Investigation, the Drug Enforcement Administration, and the Department of Homeland Security. To execute this, these superficially owned applications have to be capable to identify the officer so as to decide if he or she has the precise certificate to obtain the information. Subsequently, the information that is prone to be responsive from an intelligence and privacy outlook is ought to be secluded while in shipment. Finally, the device on which the officer collects the information should be able to storing that information strongly [13]. In our previous works, we have anticipated efficient and secure information sharing protocols for secure exchange of confidential information amongst government intelligence agencies [19, 20, 42]. This paper is superior version of our prior research paper [42].

In this paper, we present intangible outline of vertical integration, semantic interoperability architecture and framework as well as a well-organized and secluded information sharing approach for security personnels to share confidential information among them and with government departments which pact with security. This paper portrays various interoperability framework solutions. The anticipated security approach is mainly modified to fit in the following circumstances. Deem, as, a restricted law enforcement officer at a normal traffic stop. The customary protocol for traffic control demands the officer to appeal and prove the individual's driving license and vehicle registration. Still, the law enforcement officer could also wish to ensure with a wide range of other computer applications, such as migration databases, criminal information and intelligence repositories, and counter-drug intelligence databases that may be owned by external organizations, such as Central Bureau of Investigation (CBI), the Drug Enforcement Administration, and the Department of Homeland Security. The correctness and the amount of information shared stuck between security personnel and communicating government intelligence departments is rooted in the predefined grade of the security personnels. The proposed function and cooperation based information sharing approach attains data integrity using a cryptographic hash function, MD5 Algorithm; confidentiality and authentication using Public Key Infrastructure (PKI) and department confirmation using a unique and complex mapping function.

The vital outline of the paper is as follows: An undersized review of some current researches allied to the proposed approach is given in Section 2. Conceptual overview of semantic interoperability architecture and interoperability frame work solution is given in Section 3. The proposed role and cooperation based information sharing approach for security personals are presented in Section 4. The experimental results are presented in Section 5. Lastly, the conclusions are summed up in Section 6.

## 2. REVIEW OF RELATED RESEARCH

Copious researchers have offered approaches for secure and effective information sharing between communicating parties. Among them, a few researchers have offered approaches for securely sharing confidential information among government departments. Newly, rising resourceful approaches for firmly sharing secret information among government agencies and departments has drawn much attention. A concise review of some contemporary researches is presented here. To deal with the information sharing concern amid government agencies, Peng Liu *et al.* [18] have offered an inventive interest-based trust model and an information sharing protocol, which is integrated a group of information sharing policies also information exchange and trust cooperation are interleaved and equally dependent on each other. Additionally, the up-and-coming technology of XML Web Services was utilized during the accomplishment of the proposed protocol. The accomplishment was utterly unfailing with the Federal Enterprise Architecture reference models and can be unambiguously incorporated within recent electronic government systems. Jing Fan *et al.* [21] have anticipated a theoretical model for information sharing in an electronic government road and rail network. They established that the Government-to-Government (G2G) information sharing model will aid in giving knowledge for G2G information sharing and help decision makers in formulating decisions concerning the involvement in G2G information sharing. The proposed conceptual model was tartan to find out the aspects influencing the participation in an electronic government information sharing and underscore the conceptual model via case study beneath Chinese government system. Fillia Makedon *et al.* [10] have presented a negotiation-based sharing system called SCENS: Secure Content Exchange Negotiation System which was being constructed at Dartmouth College with the aid of plentiful interdisciplinary skilled. SCENS was a multilayer scalable system that guarantees transaction safety via a number of security mechanisms. It was based on the metadata description of assorted information which is appropriate to various diverse domains. They represented that with sensitive and distributed information the government users can accomplish settlement on the conditions of sharing through negotiation.

Xin Lu. [22] have established a dispersed information sharing model as well as inspected the technique standard support of the model. It was presumed that the expenditure of dealing with government information exchange and cooperation between agencies will be reduced by a raise in the prospective and efficiency of agencies' collaboration down to the secure e-government information sharing elucidations. Nabil R. Adam *et al.* [23] have inspected the demands in integration, aggregation and secure sharing information to aid situation awareness and response at the premeditated level. On removal of data from various independent systems, the system filters, integrates, and proficiently envisages information crucial to obtain a general operational picture, by utilizing context-sensitive parameters. One substantial demand was to assist secure information sharing. Sharing of information protracts to be a major complexity due to the data privacy and ownership concerns as well as owing to a widespread range of security policies followed inside various government agencies. Nabil Adam *et al.* [16] have offered a two tier RBAC approach to facilitate security and discriminative information sharing amid virtual multi-agency response team (VMART) as well as when there is require, it allows VMART expansion by permitting new collaborators (government agencies or NGOs). They also presented a coordinator web service for every member agency. The coordinator web service captures the responsibilities such as, authentication, information broadcasting, information acquisition, responsibility creation and enforcement of predefined access control policies. Awareness of secure, selective and fine-grained information sharing was skilled by the encryption of XML documents in par with analogous XML schema defined RBAC policies.

Achille Fokoue *et al.* [24] have established logic for risk optimized information sharing through rich security metadata and semantic knowledge-base that detains domain specific concepts and relationships. They long-established that the method was: (i) flexible: e.g., tactical information decomposing sensitivity in agreement with space, time and external events, (ii) situation-aware: for example, encodes need-to-know based access control policies, and further outstandingly (iii) supports elucidations for non-shareability; these elucidations along with rich security metadata and domain ontology allows a sender to intelligently execute transformation of information with the goal of sharing the transformed information with the recipient. Additionally, they have explained a secure information sharing architecture with the help of a universally accessible hybrid semantic reasoner as well as showed a number of descriptive cases that highlights the benefits of the method while complementary with conventional methodologies. Ravi Sandhu *et al.* [25] have presented a way to share secure information easily through modern Trusted Computing (TC) technologies which is not available with pre-TC technology. They have configured the PEI framework of policy, enforcement and implementation models, and confirmed its applicability in inspecting the issue and generating solutions for it. The structure enables the profound exploration of prospective TC applications for safe information sharing in the forthcoming effort. TC applications exclusive of information sharing as well are expected to be scrutinized. A group of policy-based technologies to present improved information sharing among government agencies without waning information security or person`s privacy has been developed by Tryg Ager *et al.* [26]. The method covers: (1) fine-grained access controls which sustain deny and filter semantics for accomplishment of complex policy conditions; (2) a oppressive policy ability that facilitates mixture of information from various resources conforming to each source's innovative exposé policies; (3) a curation organization which permits agencies to use and scheme item-level security categorizations and disclosure policies; (4) an auditing system which deals with the curation history of every information item; and (5) a origin auditing method that tracks derivations of information in surfeit of time to present support in assessments of information quality. The final idea was to facilitate a capacity to resolve amazing information sharing issues in government agencies and proffer ways for the growth of future government information systems. Gail-Joon Ahn *et al.* [27] have dealt with the problem of supporting choosy information sharing while reducing the possibility of illegal access. They have proposed system architecture by integrating a role-based delegation framework. Additionally by implementing a proof-of-concept, they have confirmed the practicability of their framework.

Mudhakar Srivatsa *et al.* [28] have presented a calculus approach for secure sharing of strategic information. Three operators: $\Gamma$, + and $\cdot$ are support by the security metadata which they have modeled as a vector half-space (as against a lattice in a MLS-like approach). A metadata vector is mapped into a time responsive scalar value by the value operator $\Gamma$. On the metadata vector space that are homomorphic, arithmetic is supported by the + and $\cdot$ operators with the semantics of information transforms. In order to compute the tightness of values estimates in the approach, they have developed real realizations of their metadata calculus that solves weak homomorphism without getting affected by metadata extension utilizing B-splines (a class of compact parametric curves). Muntaha Alawneh and Imad M. Abbadi [29] have offered a mechanism that enables the source organization to send content based on organization policy and requirements to another collaborating organization in such a manner that it could be accessed only by a exact a specific group of users performing a specific task or by all device members in the target organization. They have consummated this by providing a hardware-based origin of trust for the master organizer and organization devices making use of trusted computing technology.

## 3. CONCEPTUAL OVERVIEW OF VERTICAL INTEGRATION, SEMANTIC INTEROPERABILITY AND DIGITAL GOVERNMENT INTEROPERABILITY FRAMEWORK

### 3.1. Conceptual overview of vertical integration and digital government interoperability

#### 3.1.1. Vertical Integration

The spotlight is at present moving to renovation of government services moderately than automating and digitizing vacant processes. Building government electronic or digital is not merely a matter of putting existing government services on the Internet. What should and will be phenomenon are everlasting changes in the government processes themselves and perhaps the concepts of the government itself. While electronic commerce is redefining personal business and society in terms of processes and products, electronic or digital government proposals should be accompanied by re-conceptualization of the government service itself. In the long run, the full advantage of electronic or digital government will be recognized only when organizational changes escort technological changes. It is predictable that vertical integration within analogous functional walls but across different levels of government will happen first, because the breach between the levels of government is much fewer that the difference between functions. Many state agencies cooperate more closely with their federal and local counterparts than the other agencies in the identical level of government. In this stage level federal, state, and local complement systems are expected to connect or, as a minimum, communicate to each other. In accordance with survey of Momentum RESEARCH Group, citizens favor to access information through their local portal since they are most well-known with the services offered by the local government [30]. One application of vertical integration could be the business license application process. One instance of the vertical integration can be found on the Washington state website, in which federal employer identification number (FEIN) can be requested throughout the same process as state business license.

The goal of vertical integration is to faultlessly integrate the state's system with federal and local systems for traverse referencing and checking, and it has an effect of linking states to other states. An example would be construction of a national crime database, which includes DMV files regarding vehicle registrations and drivers' licenses, a master name index file for serious arrests, and traffic accidents. Yet, most of these systems are now law enforcement accessible only and not obtainable to the citizens. The next section of this paper describes a role and cooperation based security approach for security personnels of whichever modern digital government to provide homeland security. Communication and integration oriented technologies becomes more important in case of vertical integration. To integrate agencies in the state governments with their local government and federal government counterparts, technically, a web of remote connections is a requirement. In this far-flung connection and virtual transactions, several technological issues emerge; signal authentication, format compatibility of electronic data interchange, contact level of domestic legacy system to outside, etc. Vertical integration is not latest concept. State Universities and local school districts have worked together for years by having high school students take university levels classes. In short, vertical integration needs technical as well as semantic interoperability amongst communicating parties. A good example of semantic interoperability and vertical integration is New Zealand government education system. Albeit the vertical integration may offer enhanced efficiencies, privacy and confidentiality issues must first be considered. In proportion to a report from intergovernmental advisory board, the "leading" issue when developing such systems is ensuring the privacy of the citizen asking for service [31]. Government must provide suitable stability between the privacy of individual information and the right of individuals to right of entry the public records. The following section describes the idea of interoperability and their architectural overview.

### 3.1.2. Interoperability

Casually, and analyzing its own name, “inter-oper-ability” is an attribute referring to the capability of various systems and organizations for working together (inter-operate). Interoperability can be presented as: the capability of information and communication technology (ICT) systems and the business processes they hold up, to switch over data and to permit sharing of information and knowledge [32].

### 3.1.3. Digital government interoperability

Specifically, interoperability in digital (or electronic) government may refer to public and private sector organizations working in collaboration for delivering public services; managing networked environments for criminal activities prevention, terrorist attack prevention and disaster prevention, as well as, for citizen engagement in government decision-making processes, and so on. Digital (or electronic) government interoperability can be defined as: the capability of public authority’s information and communication technology (ICT) systems and business processes to share information and knowledge within and across organizational boundaries to better support the stipulation of public services as well as compose stronger support to public policies and democratic processes [32].

### 3.1.4. Technical Interoperability

Technical interoperability is mostly addressed by open standards at different levels such as connectivity, information access, data and application integration, and content management [32][33]. Technical interoperability involves connecting computer systems and services through the use of standards for interfaces, connectivity, data integration, middleware, data presentation and accessibility functions. Technical interoperability addresses problems logically located in different layers; from the bottom layer - responsible for the physical exchange of data, typically addressed by providing a set of suggested communication protocols and standards for data exchange; to the upper levels - responsible for providing technologies supporting organizational and semantic interoperability issues; through a set of mid-layers-responsible for various issues, such as, the transport mean, and engines for coordinating the execution of processes, among others. Subsequent, an example is introduced for technical interoperability. Such as, the interoperability framework could delineate that messages underneath these associations should be written in XML, and should be switched over using the Extensible Message Gateway[34], a software infrastructure component.

### 3.1.5. Semantic interoperability

Semantic interoperability is branch of the interoperability dare for networked electronic (or digital) government organizations. Inter-organizational information systems can merely work if they are capable of communicates and works with other such systems and interacts with people. This requisite can simply be meeting if communication standards are useful. A standards-based technology proposal permits partners to perform a conventional business function in a digitally improved method. An essential universal information systems scheme is a set of standards that permits network applicants to communicate and conduct business procedure electronically [35]. Addressing semantic interoperability entails considering [32]: 1) developing electronic (or digital) government ontologies – providing a common vocabulary for electronic (or digital) government plus for specific areas; 2) defining user-friendly metadata – providing understandable and straightforward metadata facilitating search processes in government websites; 3) maintaining semantic definitions – maintaining modern developed ontologies reducing the risk of divergence of local ontologies; 4) collecting data once – capturing information from citizens and businesses once, and reusing it for multiple purposes, previous consent of the data owner; 5) solving semantic obstacles – providing mediations for solving semantic problems, such as different labels for the same content, different formats for the same content, and different abstractions for modeling the identical area.

## 3.2. Conceptual Semantic interoperability architecture framework

Interoperability is not straightforward, and has many aspects. A fact that is also reflected in the many definitions provided:

(1) IEEE defines interoperability as*:* the ability of two or more systems or components to exchange information and to use the information that has been exchanged. (2) ISO/IEC 2382-01 defines interoperability as: the capability to communicate, execute programs, or transfer data among various functional units in a manner that requires the user to have little or no knowledge of the unique characteristics of those units. These explanations spotlight on the technical side of interoperability. It has also been pointed out that interoperability is often more of an organizational issue, including issues of ownership, people, usability and business processes. (3) Paul Millers [36] offers another definition: To be interoperable one should energetically be occupied in the unending progression of ensuring that the systems, procedures and culture of an organization are handled in such a way as to exploit opportunities for exchange and re-use of information, whether internally or externally. Interoperability can be realized at various levels, including: (a) Level 1: Technical interoperability: A communication protocol exists for swapping data among partaking systems. On this level, a communication infrastructure is recognized allowing systems to exchange bits and bytes, and the original networks and protocols are clearly defined. (b) Level 2: Syntactic interoperability: A general protocol to structure information is added. The format of the information exchange is definitely defined. For instance, a comma enclosed file exchange, or the XML syntax. (c) Level 3: Semantic interoperability: A general information exchange reference model is added. On this level, the meaning of the data is shared and clearly defined. Higher levels of interoperability may comprise pragmatic, dynamic, conceptual, legal, international interoperability.

#### 3.2.1. New Zealand Education Sector Architecture Framework

Education Sector Architecture Framework (ESAF) for New Zealand Government is the example of semantic interoperability and it covers vertical integration system. The New Zealand Education Sector contains various organizations. These organizations sprint their IT systems autonomously to execute their intention, they also team up and share a significant amount of information to make the Education Sector function all together. For example: (1) A student moves to a new school. The student's data moves to the IT system of the new school; (2) Schools send their enrolment data to the Ministry of Education; (3) New Zealand Tertiary Education Commission (TEC) shares the course register with providers and other agencies; (4) The Ministry provides the latest education provider information (5) New Zealand Qualification Authority (NZQA) receives assessment tests and returns test results. Basically, semantic interoperability is realized, as: (a) date exchange partners have a recognized common thoughtful of their mutual data, and (b) data exchanges remain to that common understanding. The Sector's stated idea for semantic interoperability is to create a sector data model that describes shared sector data so that sector participants can offer, manage access and realize the data. Semantic interoperability is vital to the Education Sector's performance.

#### 3.2.2. Need for an open standard methodology

The transformation of any semantic model into XML requires a documented and proven methodology in order to: (a) transform over and over again and traceably; (b) insist on reprocess and fulfillment; (c) design according to a reliable XML standard and (d) progress with suitable versioning in place. Developing a national semantic model, the New Zealand Education Sector Data Model (ESDM), is time-consuming. The New Zealand Education Sector has analyzed various options, and then selected a semantic interoperability solution. The solution amalgamates compatible open standards to the greatest extent possible, maximizing e-GIF fulfillment and thus interoperability in a broader sense. Customizations have been kept to a minimum. This described solution is suitable for any other sector or industry in a similar situation. The solution selected by the New Zealand Education Sector comprises several components such as: (1) Custom semantic model; (2) XML architecture; and (3) Model driven architecture. The components of New Zealand Education sector is discussed as follows: (1) Custom semantic model: In the absence of an appropriate global semantic model for Education, the New Zealand Education Sector has decided to develop its own semantic model, the New Zealand ESDM. Act appropriate to e-GIF, the Unified Modeling Language (UML) class diagram is the preferred modeling details for ESDM. ESDM currently defines over 300 classes, 900 attributes, 300 associations and 100 generalizations, and is rising. The selected design methodology and standard defines the: (a) Naming standard, this is based on ISO 11179-5; (b) allowable data types, which are based on UN/CEFACT data types. ESDM could easily be put back with either further semantic data model that complies with the above standards, making this solution very convenient. (2) XML architecture: The preferred design methodology and standard defines the XML architecture. There are two types of XML documents: (1) XML Instance Documents: These documents contain actual data; and (2) XML Schemas: These schemas define allowable XML constructs. Further, the XML Schemas are divided into: (a) XML Document Schemas: These documents schemas define allowable structure and content of a XML instance document; (b) XML Library Schema: This defines the pool of reusable XML components. The XML architecture enables reuse of XML components. In addition, it allows the methodology to restrict XML document schemas to be composed of pre-defined and thus approved library components only.

(3) Model-driven architecture: The UML semantic data model is the master source for shared and agreed understanding of the meaning of data. XML Schema models are derived from the UML master model, and used to generate XML run-time schemas, which are never modified straightforwardly. The model-driven architecture enables: (1) standard compliance checking; (2) naming compliance checking; (3) UML vs. XML consistency checking; (4) change logging; (5) usage reports; (6) Impact analysis; (7) version control; (8) XML schema code generation.

## 3.3. Conceptual Overview of Digital Government Interoperability Framework

The section of this paper introduces examples of interoperability frameworks adopted by electronic (or digital) Government leaders for addressing interoperability. In this paper, we have presented conceptual overview of interoperability frame work. The following interoperability frameworks are offered such as: (1) European Interoperability Framework (EIF); (2) New Zealand e-Government Interoperability Framework (NZ e-GIF); (3) The Hong Kong SARG Interoperability Framework, and (4) e-Government Interoperability Framework (e-GIF). For each of them, the following items are depicted: (a) name; (b) source i.e. person or organization accountable for its publication; (c) solution type ; (d) aim; (e) description; (f) process i.e. whether the solution includes a process for managing its content; and (vii) interoperability support – concise evaluation of the support provided by the solution to technical, organizational and semantic interoperability.

### 3.3.1. European Interoperability Framework (EIF)

**Name:** European Interoperability Framework (EIF)
**Source:** EIF was published by the Interoperable Delivery of European e-Government Services to public Administrations, Businesses and Citizens [37] [2]. IDABC is a society program run by the European Commission's Directorate-General for Informatics.
**Solution Type:** Interoperability Framework
**Aim:** EIF focuses on enhancement, rather than substitute, national interoperability guidance by adding together the pan-European dimension. It defines a set of recommendations and guidelines for electronic (or digital government) government services, in order that public administrations, enterprises and citizens can cooperate across borders in a pan-European perspective.
**Description:** EIF defines generic standards and provides recommendations on all three types of interoperability for European e-Government [38]. The framework defines three interaction types in the general form of interoperability: (i) direct interaction between citizens or enterprises of a member state with administrations of other member states and/or European institutions; (ii) exchange of data between administrations of different member states; and (iii) exchange of data between various European Union (EU) Institutions/Agencies, or between an EU Institution/Agency and one or more administrations of the member states. Tangible offers are provided for dealing with the three phases of interoperability within the EU. Recommendations for technical interoperability includes: 1) a common guideline to be based on open standards; 2) front-office technical interoperability to include data presentation and exchange, multi-channel access, file types, document formats and character sets; 3) back-office technical interoperability to include EDI- and XML-based standards, Web Services, data integration and middleware, services for message-storage, message transport and security, network services, directory and domain name services, distributed application architecture, and mailbox access; 4) guidelines for technical multilingualism, including machine translation software, facilities for citizens and endeavors to submit requests in their own language when achievable, and a totally multi-lingual top-level EU portal interface.
**Recommendations for organizational interoperability comprise:** 1) recognition and prioritization of services provided at pan-European level to be together indomitable by participate administrations via demand-driven approach; 2) agreement on the necessary business interoperability interfaces (BII) through which business processes of public administrations will be able to interoperate; 3) formalization of the expectations of several public administrations contributing to the proviso of a pan-European electronic (or digital) government service.
Recommendations for semantic interoperability include: 1) publication of information on the data elements to be exchanged at the national level and agreement on the data dictionary and multilateral mapping tables based on center pan-European electronic government data elements; 2) Pan-European synchronization of linguistic traces of specific legal vocabularies used in delivering services; and 3) development of common semantics based on XML vocabularies and considering the agreed core electronic government data elements, and the provision of specific European schemas and definitions through common infrastructures.
**Process:** The Framework does not offer progression support.
**Interoperability Support:**
1) Technical – EIF provides support for technical interoperability. For instance, the technical recommendations depicted above. 2) Organizational – EIF provides support for organizational interoperability, such as, the organizational recommendations explained above. 3) Semantic – EIF identifies foremost semantic interoperability issues and provides guidelines for addressing them at the Pan-European level. It covers up areas such as conformity on data dictionaries connected to pan-European services, and approval of common semantics as basis for XML terminology. Additionally, EIF supporting documents, such as the semantic interoperability strategy describing semantic assets - such as dictionaries, multi-lingual thesauri, cross-references and mapping tables, ontologies and services; provide both guidelines and strategies for planning and implementing semantic interoperability to support pan-European electronic (or digital) government services.

### 3.3.2. New Zealand e-Government Interoperability Framework (NZ e-GIF)

**Name:** New Zealand e-Government Interoperability Framework (NZ e-GIF)
**Source:** State Services Commission, Government of New Zealand
**Solution Type:** Interoperability Framework
**Aim:** NZ e-GIF offers a technological structure based on a covered model for organizing IT standards. The fundamental principle is straightforwardness, achieved by separation of functions into levels.
**Description:** NZ e-GIF [39] documentation consists of three documents: (i) Standards -focuses on the standards defined by GIF; (ii) Policy – explain the policies behind electronic (or digital) government interoperability framework and its development; and (iii) Resources – contains resources related to electronic-GIF. NZ e-GIF encompasses four structural layers such as: network, data integration, business services, and access and presentation. Applying to all these layers, it delineates: security, best practice, electronic (or digital) government services and web services components. Additionally, underneath all these layers it includes two more components such as: management and governance.

**1) Network** – standards for data transport. For example, standards are provided for network protocols, directory protocols, file transfer protocols, mail transfer protocols, and others. A subset of the broadly used Internet protocol suite is used.
**2) Data Integration** – standards for enabling data exchange between heterogeneous systems and data analysis on receiving systems. For instance, standards are provided for character sets, structured data, file compression, file archiving, etc.
**3) Business Services** – standards for specifying how data is mapped into usable business information and hence assigned meaning. For instance, standards are provided for discovery of meta-data, namespaces, name and address, customer relationship, directory services, digitization, statistical data and metadata, e-learning, directory services, business reporting, etc.
**4)Access and Presentation** - principles and rules casing the right to use and awarding of business systems. For example, standards are provided for website presentation, web design and maintenance, forms and authentication.
**5) Security** – standards at various levels reflecting the idea that security needs to be designed into the system, and not added as a layer on top. For example, standards are provided for data integration, web services, business services, public key infrastructure (PKI), among others.
**6) Best Practice** – standards published in this component do not ensure interoperability; they offer an approach for managing and understanding the context of information exchange.
**7) Electronic (or digital) government Services** – infrastructure components provided by the

vital coordination office for their use by government agencies.
**8) Web Services** – set of uniform purposes to mix web-based applications. For instance, standards are provided for discovery, description, access, messaging, security, and compliance.
**Process**: A process unfolding life-cycle of standards is integrated.

**Interoperability Support:** (1) Technical – The framework provides support for technical interoperability. For instance, the standards provided at the network and data integration layer and those included in the security and web services component. (2) Organizational–No support for organizational interoperability is provided. (3) Semantic – Partial support is provided for semantic interoperability.

### 3.3.3. The Hong Kong SARG Interoperability Framework

**Name:** The HKSARG Interoperability Framework
**Source:** The HKSARG Interoperability Framework was published by the Interoperability Framework Coordination Group, Government of Hong Kong SAR.
**Solution Type:** Interoperability Framework
**Aim:** The HKSARG Interoperability Framework supports the government's strategy of providing client-centric-joined-up services by making possible technical interoperability for government to government and government to public interactions [40].
**Description:** The HKSARG Interoperability Framework offers a technical specification to enable conversational interaction amongst government applications in a message-based, open environment. The technical standards proposed by the framework are grouped into high-level categories referred as interoperability domains. Under these domains, several interoperability areas are defined. A description of the domains and their relevant areas are follows: (1) Application Integration – specifications enabling application-to-application integration. The following areas are defined such as: simple functional integration in an open environment; reliable message exchange between application systems in and open environment for business document-oriented collaboration; and secure exchange of messages within web services environment. (2) Information Access and exchange – provision for file exchange, character sets and encoding. Some of the domain areas include: hypertext web content; client-side scripting; mobile web content; e-mail format; e-mail security; audio and video streaming; and document file type for content publishing. (3) Security – specifications enabling the secure exchange of information. Some of the domain areas include: IP network-level security; transport-level security; symmetric and asymmetric encryption algorithms; and digital signature algorithms, among others. (4) Interconnection – specifications enabling communication between systems. The province regions encompass: e-mail transport; mail box access; hypertext transfer protocol; directory name service; domain name service; file transfer; LAN/WAN networking; LAN/WAN transport protocol; wireless LAN; wireless LAN security; and mobile device Internet access. Moreover, the HKSARG Interoperability Framework describes the government network architecture (GNA). GNA specifies the organization and relationships between components of the government's IT infrastructure. The infrastructure components include: departmental networks (DNs), commons services (CSs), external access gateways (EAGs), and the government backbone network (GNET). GNA is depicted in **Figure 1.**

**Process:** The Framework does not offer process support
**Interoperability Support:** (1) Technical – The framework provides support for technical interoperability, basically around a set of core standards, such as: XML, SOAP and WSDL. (2) Organizational – The framework does not provide support for organizational interoperability; yet, it is thought-out to be integrated in original releases. Examples of topics to be added are: homogeny of intra-government workflow and business process management, and two categories of business processes such as public and private processes. (3) Semantic – The framework does not grant hold up for semantic interoperability. Standards for contents and resource description languages are measured for prospect embracing. In rising future semantic interoperability support, the experiences of the United Kingdom, Australia, and New Zealand Governments will be well thought-out.

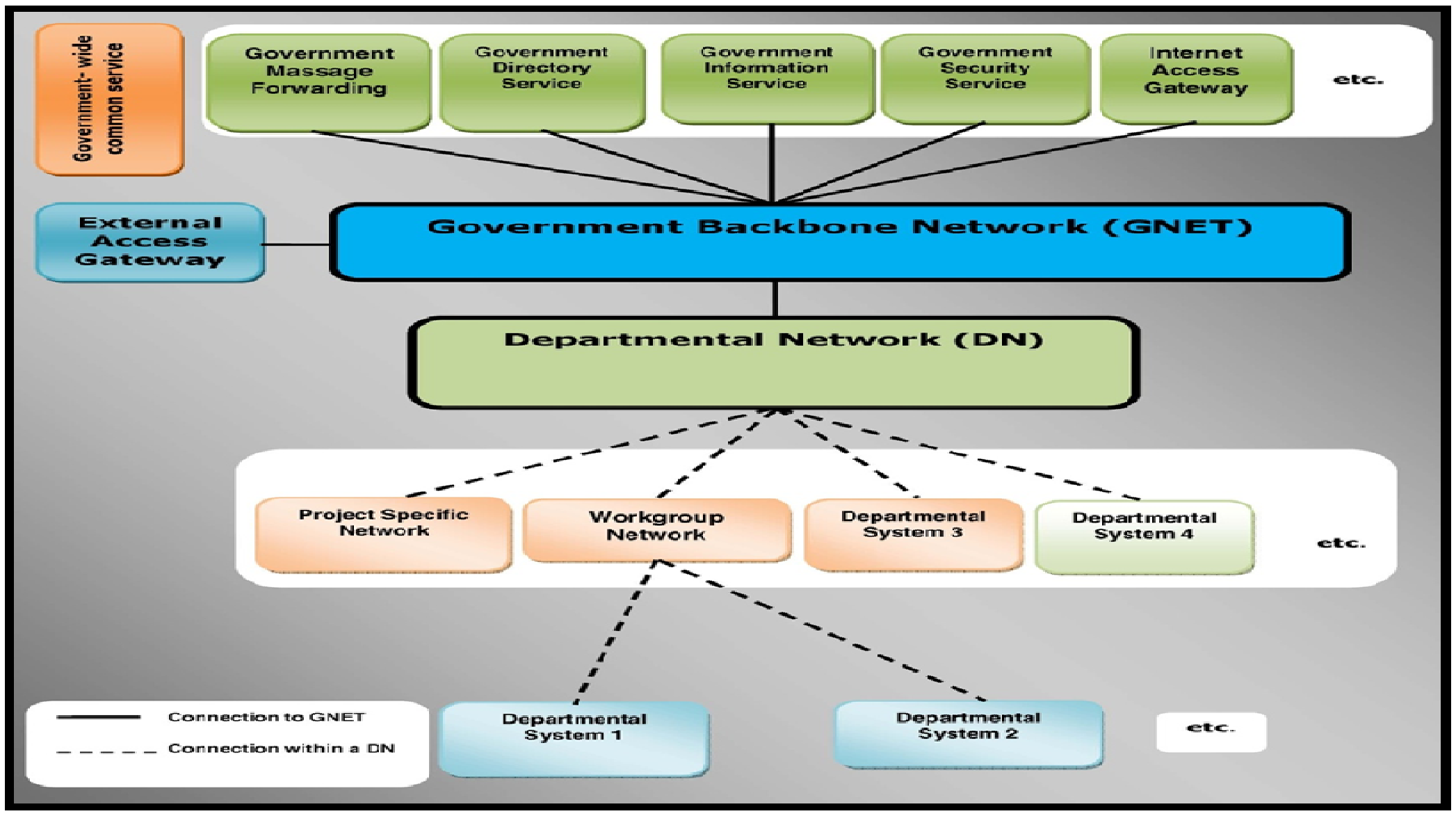

**Figure 1**: HKSARG Interoperability Framework- Government Network Architecture
Source: http://www.ogcio.gov.hk/eng/infra/download/s18.pdf

### 3.3.4. E-Government Interoperability Framework (e-GIF)

**Name**: Electronic Government Interoperability Framework (e-GIF)
**Source:** e-GIF was published by e-Government Unit, Cabinet Office, Government of United Kingdom. Electronic-GIF first release was published in September 2001.
**Solution Type:** Interoperability Framework
**Aim:** The aim of e-GIF is to allow the faultless flow of information diagonally government or public service organizations.
**Description:** e-GIF is structured in two main components [41]:
(1) Electronic (or digital) government interoperability framework (e-GIF) covers up- high-level policy statements, technical policies and management, and rules for accomplishment and execution. (2) Electronic-GIF Registry – covers up the electronic government metadata standards (e-GMS), and government category list (GCL), the government data standards catalogue (GDSC), XML schemas and the technical standards catalogue (TSC).
**Process:** e-GIF defines stakeholders' farm duties and functions for executing and preserves the framework. Additionally, it defines governance and working structures for running and developing interoperability-related issues. Some exemplar of these structures include: (a) senior information technology forum – responsible for addressing joint issues allied with possession and implementation of government IT projects; (b) interoperability working group – responsible for e-GIF definition and maintenance; (c) government schema group –responsible for setting the specifications for and harmonize the production of XML schemas for use across the public sector; and others.

**Interoperability Support:** (1)Technical – The framework supports technical interoperability, by defining a set of technical policies and specifications governing information flows across government and the public sector in general. (2) Organizational – The framework offers controlled support to organizational interoperability. Such as, it provides plan for multi-channel release of public services. (3) Semantic – e-GIF in some measure supports semantic interoperability. Such as, it offers the electronic government metadata standard (e-GMS) which indicates how public sector bodies in the United Kingdom should label content, for example, web pages and documents so that make such information easily manage, discoverable and shared. It also provides guides for semantic specification of electronic data interchange and messaging services. To finish, we can able to comprehend that vertical integration desires semantic and technical interoperability among communicating entity. Semantic interoperability also requires technical interoperability. In a nutshell, semantic interoperability is the capability of software systems (for instance, server) to make sure a consistent (or upright) understanding of shared information across numerous organizations (government or private). This outline of interoperability permits systems to significantly exchange and use information received from wide-ranging sources based on pre-defined agreements (such as, predefined trust or predefined rank) followed by the communicating parties (such as, security personnel and government intelligence department). In the following section, we have offered a role and cooperation based secure information sharing approach for security personnels to provide homeland security of every modern digital government. The proposed approach is well-matched to semantic interoperability framework and the security approach covers up vertical integration method. The details implementation of this security approach is talked about in the following Section 4.

# 4. SECURE INFORMATION SHARING SECURITY APPROACH FOR SECURITY PERSONNELS

Governments must keep in trust the critical asset, government information and manage it effectively. A greater priority must be given by government organizations at all levels for the exchange of information and data between and amidst its trusted partners. Information must be leveraged and assisted by coordinated and integrated solutions so as to meet the increasing needs and service requirements. The current "stove piped" environment has hindered the information sharing or exchanges among the agencies, the central government and the local jurisdictions. The lack of common data vocabularies for government intelligence departments has made information sharing with them both costly and complicated. Despite the fact that some improvement has been made, to specify how information sharing responsibilities and relationships, including proper central incentives will advance this task, more endeavors are needed [6]. Building secure information sharing mechanisms for security personnels is not trivial because security personnels worry that their interests may be jeopardized when they share information with government departments that are dealing with security [13]. The primary motivation behind this research is the design of well-organized and secure information sharing security approach for securely exchanging confidential and top secret information among security personnels and government intelligence departments. Although the proposed approach is non-privacy-preserving, it assures paramount confidentiality and authentication in information transfer for both the security personnel and the target government departments. In general, the security personnels obtain secret information about suspicious persons and their activities from the government intelligence agencies. During the exchange, if the information is hacked by somebody, the security personnel's further actions will go wrong, which leads to a critical issue. This demands an efficient and secure approach that offers confidential and authenticated information sharing without creating any issues and problems to security. Furthermore, there is a chance that the target government department may provide complete confidential information about a person to all the security personnels, which would affect the privacy of that person and leads to information leakage. The above case cannot be entirely averted in a non privacy-preserving approach but could be controlled by permitting information transfer based on the security personnel's rank. In the presented approach, the credibility of information shared is based on the grade or rank of the security personnel. A master control is established in the proposed security approach to monitor and control the information exchange between the security personnel and the government intelligence departments. The proposed secure information sharing security approach requires the following: a) The public key of the security personnel, the master control and the communicating departments b) A unique and complex mapping function to uniquely identify the security personnels, the master control and the communicating intelligence department. The security personnels, the master control and the communicating government intelligence departments attain their public and private keys from a trusted Certificate Authority (CA). The chief steps involved in the proposed information sharing security approach are presented in the following sub-sections

## A. Steps in the proposed security approach at the security personnel side

### 4.1. Structuring of the security personnel's query

The security personnel sends request for some secret information about susceptible persons and their suspicious activities to the government intelligence departments. It is the duty of the security personnel to transmit the request in an unintelligible possibly encrypted manner such that the hackers cannot extract any valuable information or alter the information in the request. The structuring of the security personnel's request involves the following steps:

(1) A random number $RV$ is elected and encrypted using the security personnel's public key $K_S^{Pub}$. This encrypted random number $E_{RV}$ will be used to verify if the response corresponds to the apt security personnel's request. $E_{RV} = Enc[RV]_{k_S^{Pub}}$ .

(2) After that, a set of random values $R$ are chosen and they are combined with the encrypted random number $E_{RV}$ and the request to obtain the $SE_{Data}$. The random values set $R$ will be utilized in the validation of the identity of the target government department.

$$R = \{r_1, r_2, r_3, \cdots, r_n\} \quad SE_{Data} = [E_{RV} + R + Query]$$

(3) With the help of the MD5 Algorithm, the hash value $H_v$ is computed from the $SE_{Data}$.

$H_v = MD5[SE_{Data}]$

(4) The security personnel's request is then encrypted with the target government department's public key in order to avoid others from hacking or altering the request. As the request is encrypted with the target department's public key, it can be decrypted and viewed only by the target department. $S_{Query} = Enc[\text{Query}]_{k_R^{Pub}}$

(5) The hash value $H_v$, the set of random values $R$ and the encrypted request $S_{Query}$ are combined and encrypted with the security personnel's private key $K_S^{\Pr i}$ to obtain $SA_{Data}$. The encryption with the security personnel's private key genuinely authenticates the security personnel's request. $SA_{Data} = Enc[\text{R} + \text{S}_{\text{Query}} + H_v]_{k_S^{\Pr i}}$

(6) The encrypted random number $E_{RV}$ and the obtained $SA_{Data}$ are combined and encrypted with the public key $K_C^{Pub}$ of the master control to form the security personnel's request $S_{msg}$

$S_{msg} = Enc[E_{RV} + SA_{Data}]_{k_C^{Pub}}$

The structured security personnels' request $S_{msg}$ contains the encrypted random number $E_{RV}$, and the obtained $SA_{Data}$, all encrypted with the master control's public key $K_C^{Pub}$. Now, this structured request $S_{msg}$ is transmitted to the master control.

## B. Steps in the proposed security approach at the Master Control

### 4.2 Validation of the security personnel's request

On receiving the request from the security personnel, the master control must authenticate the security personnel followed by validating the integrity of the security personnel's request. Then, the master control will add its identity to the request and send the same to the target government department. The steps involved in the integrity checking and authentication of the security personnel's request are as follows:

1. The request from the security personnel $S_{msg}$ is first decrypted using the master control's private key $K_C^{pri}$. Since the security personnel's original request is encrypted with the public key of the target government department, it couldn't be viewed by the master control. As the private key is the secret property of the intended target, the target is assured that no one else can decrypt the request. $SC_{msg} = Dec[S_{Msg}]_{k_C^{\Pr i}}$

2. The $SC_{msg}$ obtained from the above step contains $SA_{Data}$ and $E_{RV}$. The $SA_{Data}$ is afterward decrypted by means of the public key $K_S^{Pub}$ of the security personnel. The successful decryption confirms that the request has started off from the claimed security personnel. $SC'_{msg} = Dec[SA_{Data}]_{k_S^{Pub}}$ ; $SA_{Data} = \lfloor E_{RV} + R + S_{Query} + H_v \rfloor$

The $SC'_{msg}$, encloses the set of random values $R$, the encrypted random number $E_{RV}$, the encrypted request $S_{Query}$ as well as the hash value $H_v$.

3. Then, the set of random values $R$, the encrypted request $S_{Query}$ and the hash value $H_v$ are united and encrypted using the master control's private key $K_C^{Pri}$ to obtain $C_{Req}$. $C_{Req} = Enc\lfloor R + S_{Query} + H_v \rfloor_{k_C^{Pri}}$

4. Consequently, the master control forms $C'_{Req}$ by combining the encrypted random number $E_{RV}$ and the formed $C_{Req}$ and afterward encrypting them with the public key of the target department $K_R^{Pub}$. As a final point, the formed $C'_{Req}$ will be sent to the target department. $C'_{Req} = Enc\lfloor E_{RV} + C_{Req} \rfloor_{k_R^{Pub}}$.

## C. Steps in the proposed security approach at the Target Department
## 4.3. Validation of the request by the Target Department

After receiving the security personnel's request from the master control, the target department must authenticate the master control and the security personnel followed by validating the integrity of the security personnel's request. The steps involved in the above processes are as follows:

1. The request $C'_{Req}$ received from the master control is first decrypted with the private key of the target department to obtain $R_{msg}$. The $R_{msg}$, consists of the encrypted random number $E_{RV}$ and the $C_{Req}$. $R_{msg} = Dec\lfloor C'_{Req} \rfloor_{K_R^{pri}}$ , $R_{msg} = E_{RV} + C_{Req}$

2. After that, the $C_{Req}$ is decrypted by means of the public key of the master control to achieve $R'_{msg}$. The $R'_{msg}$, contains the set of random values $R$, the encrypted request $S_{Query}$ and the hash value $H_v$. $R'_{msg} = Dec\lfloor C_{Req} \rfloor_{K_C^{Pub}}$ ; Therefore, $R'_{msg} = R + S_{Query} + H_V$

3. Then, the genuine query from the security personnel $S_{Query}$ is decrypted through the target department's private key, while $S_{Query}$ is encrypted with the public key of the target department. $R''_{msg} = Dec\lfloor S_{Query} \rfloor_{K_R^{Pri}}$

4. Subsequently, the set of random values $R$, the genuine query $R''_{msg}$ and the encrypted random number $E_{RV}$ are united and their hash value $\overline{H_V}$ is worked out with the aid of the MD5 algorithm. $\overline{H_V} = MD5[E_{RV} + R + R'_{msg}]$

5. If the hash value $\overline{H_V}$ computed from the above step and the hash value $H_V$ present in the security personnel's request are identical, it guarantees that the request has not been tampered during the transfer.

$$If\ \ H_V == \overline{H_V}\ \ then\ \ , Query\ \text{is not tampered}\ ,\ else\ \ Query\ \text{is tampered}\ ,\ end\ \text{if}$$

### 4.4. Structuring of response to the Security Personnel's query

After successful validation of the security personnel's request, the target departments will outward appearance response for the security personnel's query. The steps involved in structuring the response are as follows:

1. The target department's database is scrutinized once to achieve the rank or grade of the security personnel, from whom the request started off. The rank or grade signifies the level of security personnel, and it fixes on the credibility of the information that must be given to the security personnel.
2. The encrypted random number $E_{RV}$ in the security personnel's request will be kept as such in the response.
3. A mapping function $M_{fn}$, outstandingly defined between the communicating parties is retrieved from the target department's database. For each security personnel, there is a unique mapping function in the target department's database. Afterward, the acquired mapping function is applied on the set of random values $R$ in the security personnel's request to attain mapping value $M_{val}$. Afterward, its sine value is computed and represented as $M'_{val}$. $M_{val} = M_{fn}(R)$; $M'_{val} = Sin(M_{val})$ Where $R = \{r_1, r_2, r_3, ......, r_n\}$ and $M_{fn} = \{+, -, *, /\}$
4. Later, the target department determines the amount and credibility of confidential information to be shared with the security personnel on the basis of the security personnel's rank or grade obtained from Step 1.
5. The response corresponds to the security personnel's request; the designed mapping value and the encrypted random number $E_{RV}$ are united to form $RE_{Data}$. $RE_{Data} = [E_{RV} + M'_{val} + Answer]$
6. With the aid of the MD5 Algorithm, the hash value $H_v$ is premeditated from the $RE_{Data}$ $H_v = MD5[RE_{Data}]$
7. The response matches up to the security personnel's request is afterward encrypted by means of the public key of the security personnel $K_S^{Pub}$, in order that it can only be sighted by the security personnel. $S_{Answer} = Enc[\text{Answer}]_{k_S^{Pub}}$
8. The encrypted response $S_{Answer}$, the encrypted random number $E_{RV}$, the mapping value $M'_{val}$ and the hash value $H_v$ are combined and encrypted with the master control's public key $K_C^{Pub}$ to form $R_{\mathrm{Re}s}$. Lastly, the formed $R_{\mathrm{Re}s}$ will be sent to the master control. $R_{\mathrm{Re}s} = Enc[E_{RV} + M'_{val} + S_{Answer} + H_v]_{k_C^{Pub}}$

### D. Steps in the proposed security approach at the Master Control

### 4.5. Validation of Target Department's Response by the Master Control

On receiving response from the target department, the master control must make certain the following: 1) integrity of the target department's response 2) The response originated from the true or intended target (Authentication). The steps concerned in the above processes are as follows: 1. The target department's response $R_{\mathrm{Re}s}$ is first decrypted with the master control's private key $K_C^{\mathrm{Pr}i}$, which reveals the encrypted random number $E_{RV}$, mapping value $M'_{val}$, the encrypted response $S_{Answer}$ and the hash value $H_v$. $RC_{msg} = Dec[R_{\mathrm{Re}s}]_{k_C^{\mathrm{Pr}i}}$; $RC_{msg} = E_{RV} + M'_{val} + S_{Answer} + H_v$

2. The mapping value is recomputed at the master control region and evaluated with the mapping value present in the response to make sure that the response came from the intended target department. *If* $\mathrm{M'}_{\mathrm{val}} = \overline{\mathrm{M'}_{\mathrm{val}}}$ then

*The* target is valid and forward the response , *else Discard the* response

*end* if

3. After the validation of the intended target, the encrypted response, the encrypted random number, mapping value and the hash value are united and encrypted with the public key of the security personnel $K_S^{Pub}$ and is sent back to the security personnel. $CS_{msg} = Enc[E_{RV} + M'_{val} + S_{Answer} + H_v]_{k_S^{Pub}}$

## E. Steps in the proposed security approach at the security personnel side

## 4.6. Validation of Target Department's Response by the Security Personnel

On the reception of the response from the master control, the security personnel must make sure the following: 1) integrity of the target department's response 2) The response originated from the true or intended target (Authentication). 3) The response corresponds to the apt request of the security personnel. The steps concerned in the above processes are as follows:

1. The received response $CS_{msg}$ is first decrypted by means of the security personnel's private key $K_S^{Pri}$. $S_{Res} = Dec[CS_{msg}]_{k_S^{Pri}}$

2. The response is established for its integrity by computing the hash value and compares it by means of the hash value from the target department. $\overline{H_v} = MD5[E_{RV} + M'_{val} + Answer]$

$$If\ H_v == \overline{H_v}\ then\ information\ is\ not\ tampered$$
$$else\ information\ is\ tampered$$
$$end\ if$$

3. The encrypted random number in the target department's response is decrypted by means of the private key of the security personnel $K_S^{Pri}$ to make certain that the response is valid for the request prepared. $if\ (RV == Dec[E_{RV}]_{K_S^{Pri}})$, The response is valid, end if

4. After assessing all the parameters in the target departments' response, the security personnel think about it as a valid response from the valid target department.

The block diagram in **Figure 2** shows the steps involved in the validation of target department's response by the security personnel.

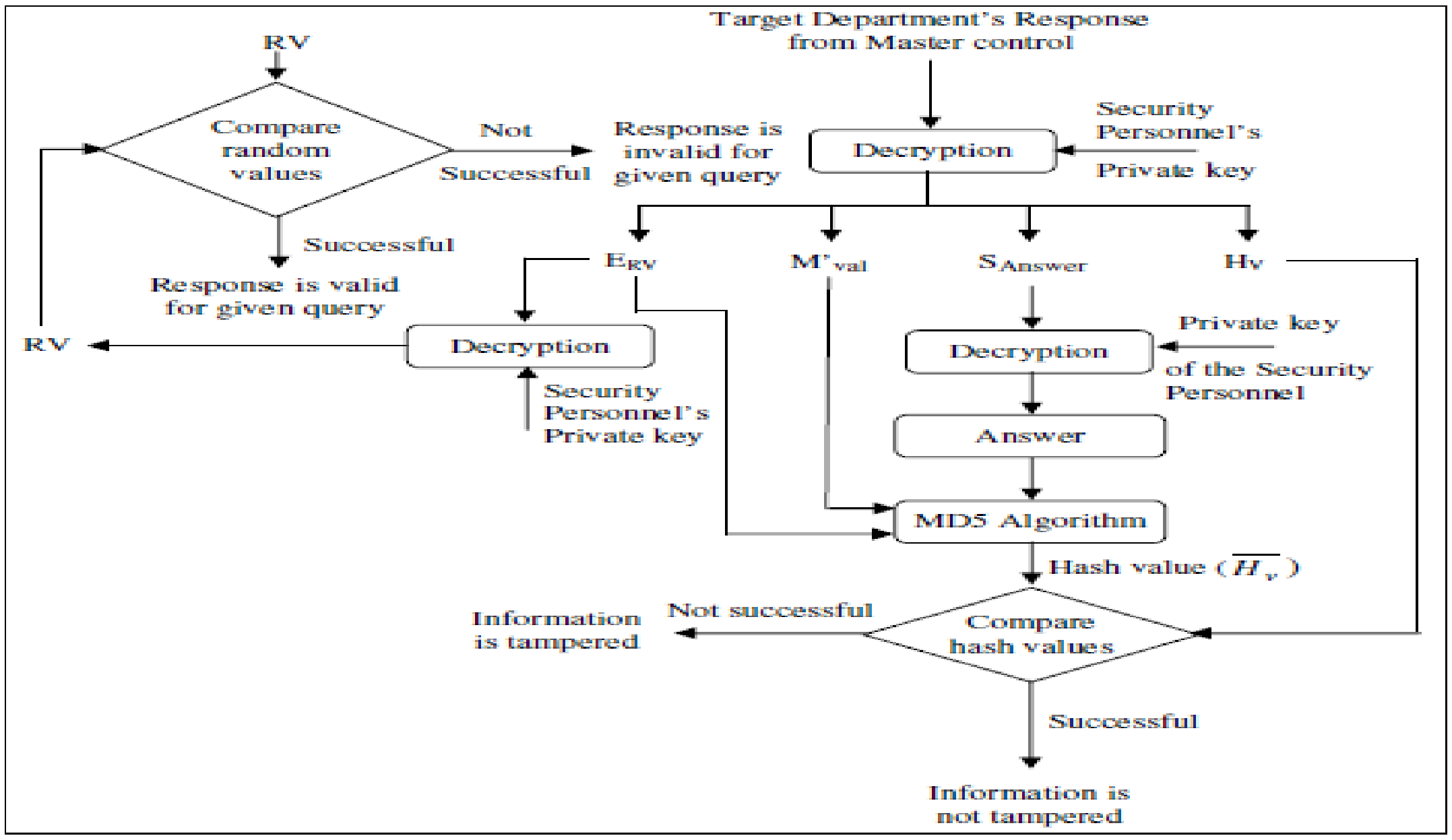

**Figure 2**: Validation of Target Department's response by the Security personnel

All the above steps guarantee that the proposed role and cooperation based approach is effective in providing confidential, authenticated and secure information sharing. Further communications between the security personnel and the government intelligence departments follow the approach discussed above.

## 5. EXPERIMENTAL RESULTS

The results obtained from the experimentation on the proposed secure information sharing security approach are presented in this section. The presented role and collaboration based information sharing security approach is programmed in Java (JDK 1.6). The results acquired from the experiments show that the presented approach provides effective and secure information sharing for security personnels and the government intelligence departments. The master control introduced in the proposed approach improves the security of information sharing by monitoring and controlling the information exchanged stuck between the security personnel and the government intelligence departments. The process started with a request for confidential information about a person, by utilizing the techniques of hashing, a unique mapping function and public key cryptography. The master control scrutinized and controlled both the request and response from the security personnel and the government intelligence department. The target department after a security verification responded with the appropriate information on the basis of the grade or rank of the security personnel. The information shared will be a compartment of the information obtainable with the target department based on the grade or rank of the security personnel.

Below, **Table 1** depicts the results obtained from the experimentation on the proposed secure information sharing security approach using reproduction data. From the Table 1, it is clear that the extent of information shared between the communicating parties depends on the rank or grade of the security personnel. In **Table 1** the field *Obtainable Information* includes the secret information about the persons and their suspicious activities, which has been composed over long periods of time and the field *rank or grade based shared Information* comprises the information shared stuck between the security personnel and the government intelligence departments. The anticipated security approach lucratively conserved the privacy of the person whose information is exchanged stuck between the communicating parties. (***See*** **Table 1**)

**TABLE 1:** RESULTS OF EXPERIMENTATION

| Security Personnel | Government Intelligence Department | Unique Identifier | Obtainable Information | Grade-based shared information |
|---|---|---|---|---|
| SP1 | Intelligence Bureau (IB) | 1 | {23,37,39,43,38, 37,24,38,35,29,40,31, 33,76,48,21,52,67, 52,71,49,26,15,38,24} | {37,24,38,35,29,40,31} |
| SP1 | Central Bureau of Investigation (CBI) | 2 | {39,33,46,56,74, 46,49,50,59, 14,6,18,29,43, 67,45,69,58,60} | {14,6,18,29,43} |
| SP1 | Narcotics Control Bureau (NCB) | 3 | {39,35,42,57,65, 49,52,64,77,87,90, 78,64,59,73,75,68, 13,17,19,24,29} | {49,52,64,77,87,90} |
| SP2 | Central Bureau of Investigation (CBI) | 1 | {19,17,36,14,23, 35,47,34,63,31,22,40, 19,12,26,18,13,17,27, 46,23,25,18,29,30} | {19, 12, 26, 18, 13, 17} |
| SP2 | Criminal Investigation Department (CID) | 3 | {62,68.65,54,57, 34,31,30,28,26, 7,16,13,27,29, 44,47,54,52,39} | {7,16,13,27,29} |
| SP3 | Criminal Investigation Department (CID) | 3 | {62,68.65,54,57, 34,31,30,28,26, 7,16,13,27,29, 44,47,54,52,39} | {44,47,54,52,39} |
| SP3 | Intelligence Bureau (IB) | 2 | {15,9,17,28,30, 85,31,17,49,27,32,46, 26,23,25,28,22,29,30, 12,7,19,13,28,31} | {12,7,19,13,28,31} |
| SP3 | Narcotics Control Bureau (NCB) | 1 | {11,26,33,15,17,45, 13,17,18,28,24,.32, 7,48,26,45,76,82, 37,21,28,17,19,25} | {13,17,18,28,24,32} |
| SP4 | Central Bureau of Investigation (CBI) | 2 | {39,33,46,56,74, 46,49,50,59, 14,6,18,29,43, 67,45,69,58,60} | {39,33,46,56,74,46,49,50,59} |
| SP4 | Intelligence Bureau (IB) | 1 | {23,37,39,43,38, 37,24,38,35,29,40,31, 33,76,48,21,52,67, 52,71,49,26,15,38,24} | {37,24,38,35,29,40,31} |

## 6. CONCLUSIONS

This paper portrays the conceptual overview and prerequisite of vertical integration, semantic interoperability and technical interoperability. This paper also depicts the details conceptual overview of interoperability framework architecture for digital government of diverse country in this world. Information giving out and integration are being contemplated as the nearly all ever more accepted methodologies by governments in the region of the world, for resolving problems in a wide assortment of programs and tactic regions. Vertical information integration necessitates semantic interoperability. In addition, semantic interoperability needs technical interoperability for speaking to winning interoperability among different department of several digital governments. Secure information giving out and integration among different system is required to improve the digital government performance. As a result information giving out and integration is the foremost issues for flourishing every digital government system. With the intention of share and integrate the information in protected manner, a role and cooperation based security approach for security personnel is offered in this article. Secured information swap over is a noteworthy attribute of every digital government that wants to promise autonomous ethics. Challenges in construction a computational infrastructure for exchanging furtive information is thorny to work out and stipulate novel spur schemes. In this paper, we have offered an efficient role and collaboration based loom for confidential sharing of secret information amid security personnels and government departments. The projected secure information sharing security approach has offered confidentiality, authentication, integrity, and agency confirmation by utilizing MD5 Algorithm, public key infrastructure and a unique and complex mapping function. As well, on the basis of a predefined grade or rank of security personnel, a restricted privacy is maintained between the security personnel and government intelligence departments. The usefulness of the proposed security approach has been recognized with the aid of experimental results. In conclusion, this proposed security approach will work in cooperation intra-organizational or inter-organizational department among security personnel and government intelligence department to offer homeland security plus to minimize the criminal activities inside the country under digital government environment. Finally, the proposed security approach will show the way to vertical information integration approach and diverse interoperability approach for every modern digital government.

## ACKNOWLEDGEMENTS

I would like to thank Dr. G.K. Pradhan, Dr. Sanjay Biswas, Dr. Puthal, Dr. Mishra, Dr. B.B. Prdhan, Dr. Ashna Anjume and my father Md. Idrish as well as mother Ashma.

## Author:

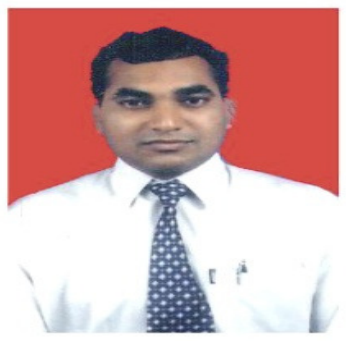

**Md.Headayetullah** received the Diploma in Computer Science & Engineering (DCSE) with 1st Class from Acharya Polytechnic, Bangalore, India and Bachelor of Engineering (B.E) degree with 1st Class from Yeshwantrao Chavan College of Engineering of Nagpur University, Nagpur, India in 2000 and 2003 respectively. He received second prize in state level for his best project in B.E degree. He received M.Tech degree with First Class with Honours from the Department of Computer Science & Engineering and Information Technology of Allahabad Agricultural Institute-Deemed University, Allahabad, India in 2005. He was the topper of the University in his M.Tech Degree. He has submitted his thesis for the award of PhD Degree in Computer Sc. & Engineering. He is working closely with Prof. (Dr.) G.K Pradhan and Prof. (Dr.) Sanjay Biswas in the Department of Computer Science and Engineering from Institute of Technical Education & Research (Faculty of Engineering) of Siksha 'O' Anusandhan University (SOAU), Bhubaneswar, India. He works in the field of E-Government, Digital Government, Networking, Internet Technology, Data Privacy, Cryptography, Information Security and Mobile Communication. He has authored more than Five Research Publication in International Journal. He is currently working as an Assistant Professor in Computer Science & Engineering and Information Technology at Dr. B.C. Roy College of Engineering, Durgapur, West Bengal University of Technology, Kolkata, India. Professor, Headayetullah is the members of IAENG and IACSIT respectively. Professor, Headayetullah is a reviewer for so many International Journals.